# Decoding 122-Type Iron-Based Superconductors: A Comprehensive Simulation of Phase Diagrams and Transition Temperatures

[1,2]Chi Ho Wong* and [1]Rolf Lortz*

[1]Department of Physics, The Hong Kong University of Science and Technology, Hong Kong

[2]Department of Industrial and Systems Engineering, The Hong Kong Polytechnic University, Hong Kong

Email(s): chkhwong@ust.hk; lortz@ust.hk

**Abstract**

Iron-based superconductors, a cornerstone of low-temperature physics, have been the subject of numerous theoretical models aimed at deciphering their complex behavior. In this study, we present a comprehensive approach that amalgamates several existing models and incorporates experimental data to simulate the superconducting phase diagrams of the principal "122-type" iron-based compounds. Our model considers a multitude of factors including the momentum dependence of the superconducting gap, spin-orbital coupling, antiferromagnetism, spin density wave, induced XY potential on the tetrahedral structure, and electron-phonon coupling. We have refined the electron-phonon scattering matrix using experimental angle-resolved photoemission spectroscopy (ARPES) data, ensuring that all electrons pertinent to iron-based superconductivity are accounted for. This innovative approach allows us to calculate theoretical critical temperature ($T_c$) values for $Ba_{1-x}K_xFe_2As_2$, $CaFe_2As_2$ and $SrFe_2As_2$ as functions of pressure. These calculated values exhibit remarkable agreement with experimental findings. Furthermore, our model predicts that $MgFe_2As_2$ remains non-superconducting irrespective of the applied pressure. Given that 122-type superconductivity at low pressure or low doping concentration has been experimentally validated, our combined model serves as a powerful predictive tool for generating superconducting phase diagrams at high pressure. This study underscores that the high transition temperatures and the precise doping and pressure dependence of iron-based superconductors are intrinsically linked to an intertwined mechanism involving a strong interplay between structural, magnetic and electronic degrees of freedom.

**Introduction**

The '122' family of iron-based superconductors, represented by $AFe_2As_2$ (where A is Ba, Sr, or Ca), has been the subject of extensive study. $BaFe_2As_2$ (Ba122), under ambient pressure, exhibits a stripe-type antiferromagnetic spin density wave (SDW) order without superconductivity. However, the introduction of external hydrostatic pressure, internal chemical pressure (e.g., isovalent doping[1,2]), or ionic substitution in systems such as $Ba_{1-x}K_xFe_2As_2$, $Ba_{1-x}Na_xFe_2As_2$, or $Ba(Fe_{1-x}Co_x)_2As_2$ can induce superconductivity. Substitutions such as replacing $Ba^{2+}$ with $K^+$ or $Na^+$ introduce holes[3], while substituting $Fe^{2+}$ with $Co^{3+}$ introduces electrons[4]. The application of pressure or doping gradually suppresses the SDW transition and gives rise to a superconducting phase. The SDW transition is accompanied by a structural change from a tetragonal high-temperature to an orthorhombic low-temperature structure[5], known as a nematic electronic transition.

Hole-doped Ba122 presents a rich phase diagram, including a re-entrant tetragonal C4 phase region that restores the fourfold symmetry in the basal plane. This phase is characterized by electron spin rearrangement and absence of electronic nematic ordering[6,7] and somewhat suppresses the superconducting transition[7,8]. $SrFe_2As_2$ and $CaFe_2As_2$ also exhibit iron-based superconductivity under pressure[9,10]. Interestingly, $MgFe_2As_2$ does not exhibit superconductivity despite magnesium's position in the 2$^{nd}$ column of the periodic table[5]. This anomaly underscores the complex interplay between structural, magnetic and electronic degrees of freedom in these materials. The coexistence of superconductivity with a momentum-dependent superconducting gap with other electronically ordered phases such as antiferromagnetic SDW and nematicity[11], and the presence of strong spin-orbital coupling further highlights the complex unconventional nature of these materials.

While it is widely accepted that antiferromagnetism enhances electron-phonon coupling on the Fermi surface in unconventional superconductors, recent studies suggest that the significance of electron-phonon coupling in unconventional superconductivity may have been underestimated. Li et al.[12] demonstrated that phonon softening in AFeAs compounds amplifies electron-phonon coupling by a factor of approximately 1.6. Deng et al.[13] interpreted out-of-plane lattice vibrations as a phonon softening phenomenon, which they incorporated into their calculations to enhance electron-phonon scattering. Coh et al. further refined these models and proposed that the electron-phonon scattering matrix in iron-based superconductors was underestimated by a factor of approximately 4[10,11]. They attributed this underestimation to the presence of nearest-neighbor interactions under an antiferromagnetic SDW, which resulted in a first amplification factor of 2, and the vertical displacement of lattice Fe caused charge transfer to induce *xy* potential in tetrahedral regions under an antiferromagnetic SDW situation, leading to another factor of approximately 2 increase in the electron-phonon scattering matrix[14]. S.-F. Wu et al.[15] also observed a significant increase in intensity for the emergent As phonon mode in the XY plane geometry.

Moreover, a discernible shift of the spectral weight between the normal and the superconducting state is evident in the photoemission spectra below the superconducting energy gap of various iron-based compounds in an energy range of approximately 30-60meV below the Fermi energy[16-19]. This shift, observed in the ARPES range, suggests that the involvement of superconducting electrons in iron-based superconductivity may have been underestimated. This underestimation could potentially account for the discrepancy between theoretical and experimental $T_c$ values based on the electron-phonon coupling method. The ARPES data[16-19] can be utilized to revise the electron-phonon scattering matrix. Given the high transition temperatures of Fe-based superconductors, it is crucial to consider the full electronic density of states (DOS) in a range of $E_F - E_{Debye}$ to $E_F$ and not only the Fermi level value, where $E_{Debye}$ represents the upper limit of the phonon energies that can be transferred to electrons. In contrast to classical low-$T_c$ superconductors, at the high transition temperatures of Fe-based superconductors, contributions of high-energy phonons in the electron-phonon scattering mechanism become significant. This approach, which is a direct consequence of energy conservation, is corroborated by ARPES experiments[16-19] where the energy range of the spectral weight shift is approximately in the order of the Debye energy.

In our quest to decode the intricate electronic phenomena in 122-type iron-based superconductors, we initially turn our attention to $Ba_{1-x}K_xFe_2As_2$. This compound, boasting the highest $T_c$ among the 122-type superconductors, presents a particularly rich phase diagram[20-23] and becomes superconducting above 0.8GPa at $x = 0$[18]. Our investigation will delve into and compare the influence of both first-order and higher-order antiferromagnetic fluctuations on the $T_c$ calculations. This comparative analysis will encompass the compounds $Ba_{1-x}K_xFe_2As_2$, $CaFe_2As_2$, and $SrFe_2As_2$. Furthermore, we will explore whether the application of pressure could potentially induce superconductivity in $MgFe_2As_2$. This comprehensive approach aims to shed light on the complex interplay of factors governing superconductivity in these materials.

**Computational algorithm**

Our preliminary $T_c$ calculations include at least 6 components.

(1) Exchange factor: The pressure dependence of the antiferromagnetic interaction can be used to monitor the variation of the exchange interaction. We define $M_{Fe}$ and $E_{co}$ as the magnetic moment of the Fe atom and the exchange-correlation energy, respectively. The exchange factor becomes $f(E_{ex}) \sim \dfrac{[M_{Fe}M_{Fe}E_{co}]_{P>0}}{[M_{Fe}M_{Fe}E_{co}]_{P=0}}$ at any external or chemical pressure $P$ [25].

(2) Coh factor: It is well accepted that the antiferromagnetic (AFM) state usually increases the electron-phonon scattering matrix by a ratio of $R_{AF}$ when the ab-initio calculations change from spin-restricted to spin-unrestricted mode. To encounter the SDW, Coh et al [14] have proposed that the antiferromagnetic SDW background in the presence of the induced *xy* potential (triggered by an abnormal out-of-plane phonon state $R_{tetra}$) in the tetrahedral region increases the electron-phonon scattering matrix of iron-based superconductors by a factor of $R_{Coh} = R_{SDW} \cdot R_{tetra} \sim 4$, hereafter referred to as the Coh factor[12], where $R_{SDW} = 2$ and $R_{tetra} \sim 2$. The SDW in the iron-based superconductors (IBSC) modifies the AFM-assisted electron-phonon coupling between the nearest neighbors, which cannot be cancelled out and instead leads to the generation of an AFM constructive interference-like pattern[14]. Whenever there is an energy transfer from AFM fluctuations to electron-phonon coupling, the maximum AFM interactions between two neighboring atoms can only combine their own AFM energy in a repeating unit. (i.e. $\delta E_{AFM} \to 2\delta E_{AFM}|_{max}$), but it turns out that the electron-phonon interaction $\lambda$ due to the AFM SDW is increased by a factor of 4 (i.e. $\lambda \to 4\lambda$), indicating a sign of a higher order AFM fluctuations with an expression of $\lambda \propto \delta E_{AFM}^2$. Using the Coh factor brings the simulation results more in line with the experimental observations[14]. However, the appearance of the induced *xy* potential requires calibrating the GGA+A functional, which is a time-consuming experimental effort and a computationally expensive mission[14]. To solve this problem, we define $R_{tetra} \sim \dfrac{0.5\left(DOS_{upper}^{XY} + DOS_{lower}^{XY}\right)}{DOS_{both}^{XY}}$ within the ARPES range, where the shift range of spectral weight occurs. $DOS_{upper}^{XY}$ represents the average electronic density of states for the structure containing only the upper tetrahedral plane, while $DOS_{lower}^{XY}$ indicates the average electronic DOS for the structure that only contains lower tetrahedral planes. $DOS_{both}^{XY}$ is the average electronic DOS representing the original structure coexisting upper and lower tetrahedral regions. The ionic interaction $V_{ion}^{XY}$ on the *XY* plane in the presence of charge fluctuations in the tetrahedral regions is $V_{ion}^{XY} \sim V_{XY} \cdot R_{tetra}$.

(3) APRES factor: To include all relevant electrons in iron-based superconductivity in the calculations, the average electron-phonon scattering matrix $g_{pp'}(E)$ within the ARPES range of spectral weight shift[16-19] is $\left\langle \sum\limits_{E_F - E_{Debye}}^{E_F} \dfrac{g_{pp'}(E')}{\varepsilon'} \right\rangle$. The dielectric constant $\varepsilon'$ controls the screening effect [25] when all the relevant electrons interact with $V_{ion}^{XY}$. Including the ARPES factor increases the electron-phonon scattering matrix by $R_{ARPES} \sim \dfrac{\left\langle \sum\limits_{E_F - E_{Debye}}^{E_F} g_{pp'}(E')/\varepsilon' \right\rangle}{g_{pp'}(E_F)/\varepsilon}$.

(4) Anisotropy factor: The ellipse equation, $p_{angular}(\theta) = \dfrac{a_{major} b_{minor}}{\sqrt{(b_{minor}^2 - a_{major}^2)\cos^2\theta + a_{major}^2}}$, is used to mimic the effect of an anisotropic momentum space when there is 4-fold symmetry (as illustrated by the two overlapping red ellipses in Figure 1). The major $a$ and minor $b$ axes control the anisotropy of the gap. The energy change due to the formation of an anisotropic momentum space is $\langle f_{angular} \rangle \sim \dfrac{8\int_0^{\pi/4} \frac{1}{2} p_{angular}(\theta)^2 d\theta}{\pi a_{major}^2}$, where $\pi a_{major}^2$ is the area of an isotropic s-wave momentum space. $\langle f_{angular} \rangle$ equals to 1 if it is an isotropic s-wave superconductor.

(5) Spin-orbital coupling SOC factor: The spin-orbital coupling of iron-based superconductors[26] is typically ~10meV, where the SOC energy can be comparable to the ARPES energy range of spectral weight shift[16-19]. The effect of SOC should be included in the calculation of the pairing-strength.

(6) Electron-phonon factor: The electron-phonon coupling is $\lambda_{PS} = 2\int \alpha_{PS}^2 \dfrac{F(\omega)}{\omega} d\omega$ where $F(\omega)$ is the phonon density of states as a function of frequency $\omega$. Taking into account the above factors, the $\alpha_{PS}^2 F(\omega)$ becomes[27]

$$\alpha_{PS}^2 F(\omega) \sim \left\langle \sum_{v_F-v_{Debye}}^{v_F} \int \dfrac{d^2 p_E}{v_E} \right\rangle \left\langle \sum_{v_F-v_{Debye}}^{v_F} \int \dfrac{d^2 p_E'}{(2\pi\hbar)^3 v_E'} \right\rangle \langle f_{angular}\rangle \sum_v \delta(\omega - \omega_{p-p'v}) \left| \sqrt{\dfrac{\hbar}{A\omega_{p-p'v}}} \int u_i \cdot \nabla(V_{XY} R_{ph}) \psi_p^* R_{SDW} R_{AF} R_{ARPES} \psi_{p'} dr \right|^2 / \left\langle \sum_{v_F-v_{Debye}}^{v_F} \int \dfrac{d^2 p_E}{v_E} \right\rangle$$

where $v_E \in (v_F - v_{Debye}, v_F)$ is the velocity in the ARPES range of spectral weight shift and $v_F$ is the Fermi velocity. The velocity $v_{Debye}$ can be interpreted from the Debye energy. $A$ is a material constant. $\hbar$ is the Planck constant over $2\pi$ and $\psi_{p'}$ is the wave function of the electrons. The $\alpha_{PS}^2 F(\omega)$ may be further reduced to $\alpha_{E_F}^2 F(\omega) \cdot R_{AF}^2 \cdot R_{SDW}^2 \cdot R_{tetra}^2 \cdot R_{APRES}^2 \cdot \langle f_{angular} \rangle$. In the case of a strong coupling, the pairing strength and the Coulomb pseudopotential are renormalized to $^*\lambda_{PS}$ and $^*\mu$, respectively[27].

The electronic and dielectric properties of the samples are computed by CASTEP at the GGA-PBE level[28-29]. The maximum SCF cycle is 100 with the tolerance of $2 \times 10^{-6}$ eV/atom. The reciprocal $k$-space interval is 0.025(1/Å). The norm-conserving potential is used for SOC calculations instead of the ultrasoft pseudopotential. The finite displacement method is used to calculate the phonon data at the LDA level, where the supercell defined by cutoff radius is 0.5nm and the interval of the dispersion is 0.04(1/Å). The exchange factor based on the mean field approach is calculated by the spin-unrestricted GGA-PW91 functional and the 1st and 2nd AFM fluctuations are investigated separately. To avoid that all simulation parameters are dependent on our computation, we use the lattice parameters and Debye temperature from the literature if available. Otherwise, the Debye temperature is taken from the CASTEP platform. While the strongly correlated electron-electron interaction is not easily calculated from the electronic DOS, Debye temperature and Fermi energy[32], the consideration of pseudopotentials in the range of 0.1 to 0.2 should make sense. All Coulomb pseudopotentials $\mu$ are set to 0.15 for fair comparison. The anisotropic factor is adjusted to demonstrate how it affects the theoretical $T_c$. The increase in electron-phonon coupling resulting from the exchange enhancement can be represented as a separable variable[30]. The separable variable representing the increase in electron-phonon coupling due to exchange enhancement, can be obtained by multiplying the electron-phonon coupling by the exchange enhancement factor[30]. It should be noted that this is not restricted to the first-order exchange interaction[30]. The parameters influencing AMF under pressure can be described by $R_{AF}^2|_{P>0} \sim R_{AF}^2|_{P=0} \cdot f(E_{ex})^2$ and $R_{tetra}^2|_{P>0} \sim R_{tetra}^2|_{P=0} \cdot f(E_{ex})^2$ under the 2nd AFM fluctuation. Similarly, we set $R_{AF}^2|_{P>0} \sim R_{AF}^2|_{P=0} \cdot f(E_{ex})$ and $R_{tetra}^2|_{P>0} \sim R_{tetra}^2|_{P=0} \cdot f(E_{ex})$ under the 1st AFM fluctuation. It is not necessary to switch on the spin-unrestricted mode to calculate $R_{tetra}^2|_{P=0}$, as we have manually removed one of the tetrahedral planes to mimic the out-of-plane phonon that appears at the spin-unrestricted GGA+A level[14]. The pairing strength is substituted into the McMillian $T_c$ formula[27]. Only Fe and As atoms are imported in the ab-initio calculation to showcase the bare pairing strength.

**Results**

Fig. 1a displays our theoretical $T_c$ of BaFe$_2$As$_2$ under external pressure compared to experimental literature data[20]. The combined model based on the 2nd AFM fluctuation shows a reasonable accuracy in the superconducting phase diagram simulation. The enhanced electron-phonon coupling, and the exchange factor are optimized at 1.3GPa, but they are drastically reduced at higher pressures, as shown in Fig. 1b. The Debye temperatures of uncompressed BaFe$_2$As$_2$ at the low and high temperature limits are 379K and 470K, respectively[31]. Our computed $\lambda_{E_F}$ of BaFe$_2$As$_2$ at 0.8GPa only increases from 0.33 to 0.37 after activation of the spin-orbit coupling. However, the Coh factor and ARPES factors are the main ingredients to increase the pairing strength to ~0.9, allowing the theoretical $T_c$ to occur above 30K. In contrast, the combined model makes use of 1st AFM fluctuation results in a significant discrepancy between the calculated and experimental $T_c$ at high pressures.

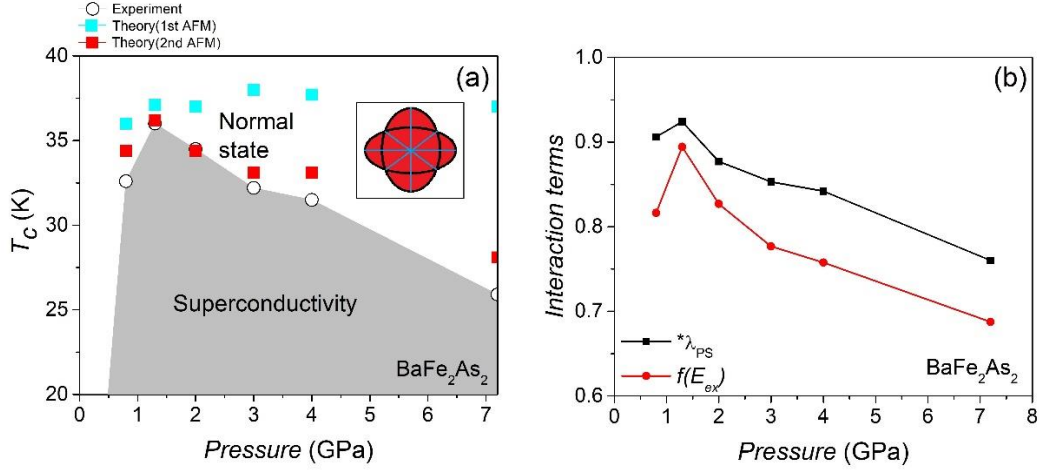

**Fig 1.** The $T_c$ distribution and the corresponding interaction terms of $BaFe_2As_2$ as a function of hydrostatic pressure. (a) Theoretical $T_c$ of $BaFe_2As_2$ under compression (red and blue squares) together with experimental data (open circles) [20]. The 4-fold symmetry is outlined by two overlapped red ellipses in which blue lines are used to split it into 8 regions in equal partition, where the area of each region is $\int_0^{\pi/4} \frac{1}{2} p_{angular}(\theta)^2 d\theta$. (b) The effect of pressure on the interaction terms. The renormalized $^*\lambda_{PS}$ is calculated under 2$^{nd}$ AFM fluctuations.

The experimental and theoretical $T_c$ values of $Ba_{1-x}K_xFe_2As_2$ are in close but not perfect agreement, as shown in Fig 2a[8]. The inclusion of either the 1$^{st}$ or 2$^{nd}$ AFM fluctuation does not have a significant effect on the calculated $T_c$ of $Ba_{1-x}K_xFe_2As_2$. We import the full set of lattice parameters measured by Rotter *et al.* as a function of doping [32]. In the range of $0.25 < x < 0.27$, the Fe moments are aligned in the out-of-plane direction[8], leading us to set a fixed orientation of the Fe moment in the out-of-plane direction in the ab-initio simulation. If the fixed orientation of the out-of-plane Fe moment is not considered, the theoretical $T_c$ could be around 32K. However, due to the out-of-plane Fe moment causing a 15% reduction in the exchange factor, the theoretical $T_c$ values are ultimately reduced by ~4K. The $Ba_{0.6}K_{0.4}Fe_2As_2$ has a highest theoretical $T_c$ of ~34K. Fig 2b illustrates the individual components of the pairing strength. The $^*\lambda_{PS}$ reaches a maximum at $x \sim 0.4$, where our computed value of $\lambda_{E_F}$ is 0.88 and the literature value is 0.9[33]. When the doping concentration is increased from 0.4 to 0.6, the $^*\lambda_{PS}$ decreases slightly. The pressure effect on the dielectric constant of $BaFe_2As_2$ and $Ba_{1-x}K_xFe_2As_2$ under spin-unrestricted conditions is small.

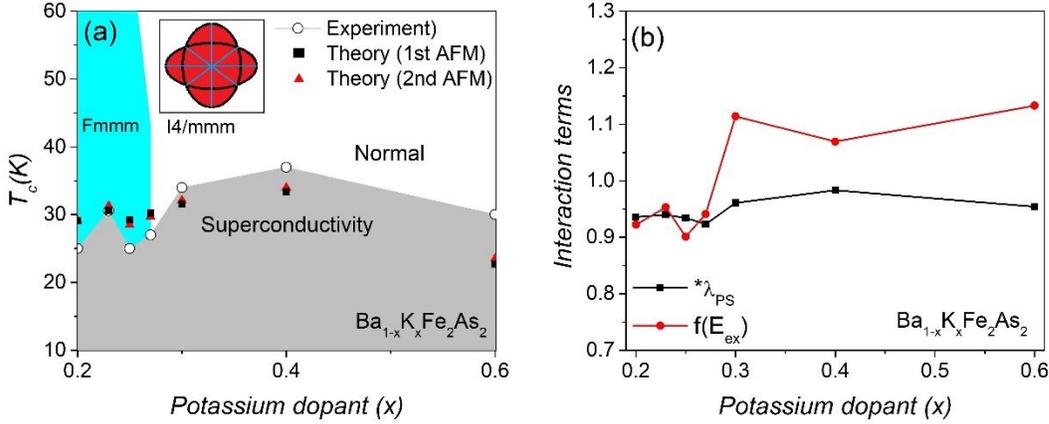

**Fig 2.** (a) The doping dependence of the theoretical and experimental [8,32] $T_c$ of $Ba_{1-x}K_xFe_2As_2$. The theoretical $T_c$ based on the 1st and 2nd AFM fluctuations are marked by squares and triangles, respectively. (b) The individual interaction terms of $Ba_{1-x}K_xFe_2As_2$ as a function of K content. The renormalized $^*\lambda_{PS}$ is calculated under 2nd AFM fluctuations.

The $T_c$ calculation based on the 2nd AFM fluctuation of the $Ba_{1-x}K_xFe_2As_2$ at $x = 0.4$ is demonstrated below. The calculated $\lambda_{E_F}$, $\langle f_{angular} \rangle$, $R_{AF}$, $R_{SDW}$, $R_{tetra}$ and $R_{ARPES}$ are 0.88, 0.87, 1.18, 2, 1.68, and 2.06 respectively. The amplified electron-phonon coupling is

$$^*\lambda_{PS} = \frac{\lambda_{PS}}{\lambda_{PS}+1} = \frac{0.88 \cdot 0.87 \cdot 1.18^2 \cdot 2^2 \cdot 1.68^2 \cdot 2.06^2}{0.88 \cdot 0.87 \cdot 1.18^2 \cdot 2^2 \cdot 1.68^2 \cdot 2.06^2 + 1} = 0.982$$ and the Coulomb pseudopotential is

$$^*\mu = \frac{\mu}{\lambda_{PS}+1} = \frac{0.15}{1 + 0.88 \cdot 0.87 \cdot 1.18^2 \cdot 2^2 \cdot 1.68^2 \cdot 2.06^2} = 0.0026.$$ Doping is also associated with an internal chemical pressure and causes a variation in the antiferromagnetic exchange by $f(E_{ex}) \sim \frac{[M_{Fe}^2 E_{co}]_{P>0}}{[M_{Fe}^2 E_{co}]_{P=0}} = \left(\frac{1.48^2}{1.41^2}\right) 1.027 = 1.132$ by which the Fe moments are scaled to the Bohr magneton. We substitute the pairing strength into the McMillian $T_c$ formula,

$$T_c = \frac{T_D}{1.45} \exp\left(\frac{-1.04(1+\lambda)}{\lambda - \mu(1+0.62\lambda)}\right) = \frac{393}{1.45} \exp\left(\frac{-1.04(1+0.982)}{0.982 - (0.0025)(1+0.62 \cdot 0.982)}\right) = 34.2K$$

By performing calculations, we estimate the $T_c$ of $SrFe_2As_2$ under compression. The theoretical $T_c$ values of $SrFe_2As_2$ at low pressures are in good agreement with the experimental data[9] presented in Table 1, regardless of whether the first or second AFM fluctuation is used. However, significant discrepancies between the theoretical and experimental $T_c$ values are observed for $SrFe_2As_2$ under high pressure conditions[9]. On the other hand, the theoretical $T_c$ value of CaFe$_2$As$_2$ as a function of pressure[10] does not vary significantly when either the first or second-order AFM fluctuation method is used. Furthermore, our spin-unrestricted calculation reveals that the magnetic moment of MgFe$_2$As$_2$ remains at zero when the pressure exceeds 3GPa, as indicated in Table 2. In addition to these results, we also investigated the relationship between the momentum space and the superconducting gap, as depicted in Figure 3. It is worth noting that the calculated values of $T_c$ exhibit minimal error ($\delta T_c \sim$ 2-4K), regardless of the presence of gap anisotropy. The theoretical $T_c$ towards an isotropic superconducting gap is slightly higher.

Table 1: Comparison between experimental[9,10] and theoretical $T_c$ of SrFe$_2$As$_2$ and CaFe$_2$As$_2$ under compression. The Debye temperature, the tetrahedral angle, lattice parameters are obtained from the literature[9,31,34,35] in the supplementary materials.

|  | Experimental $T_c$ | Theoretical $T_c$ (2$^{nd}$ AFM) | Theoretical $T_c$ (1$^{st}$ AFM) |
| --- | --- | --- | --- |
| SrFe2As2 (3GPa) | 30K | 28.3K | 30.2K |
| SrFe2As2 (5GPa) | 19K | 23.5K | 29.4K |
| CaFe2As2 (0.1GPa) | 12-15K | 14.6K | 14.6K |
| CaFe2As2 (1.2GPa) | 12-15K | 15.1K | 14.9K |

Table 2: Magnetic analysis of MgFe$_2$As$_2$ under pressure.

| P(GPa) | Fe moment ($\mu_B$) | P(GPa) | Fe moment ($\mu_B$) |
| --- | --- | --- | --- |
| 0 | 0.554 | 6GPa | 0.000 |
| 3 | 0.001 | 9GPa | 0.000 |

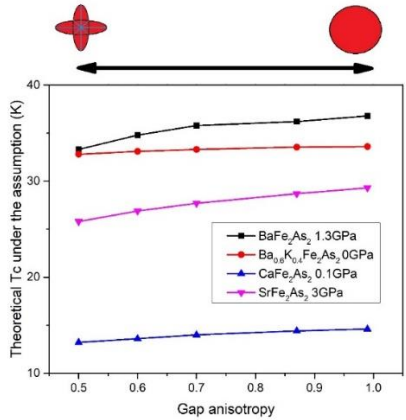

Figure 3: Effect of the momentum dependence of the superconducting gap on the theoretical $T_c$ in the studied compounds under the 2$^{nd}$ AFM fluctuation.

**Discussion**

While the here presented combined model has demonstrated efficacy in bridging the gap between theoretical and experimental $T_c$, a comprehensive theory of IBSC remains elusive. This paper does not aim to reevaluate the six models outlined in the methodology section, as their scientific validity has been established in peer-reviewed literature. We do not seek to validate the theory of iron-based superconductivity in this work. Instead, our objective is to amalgamate $T_c$ calculations from these validated studies and systematically integrate each sub-model. This approach allows us to examine its potential for predicting the theoretical superconducting phase diagram. Our focus in this section is primarily on data analysis, rather than conjecture about the triggers of iron-based superconductivity. If a proposed model of iron-based superconductors is deemed incorrect, a universal theory of iron-based superconductors would need to be already in place. However, at present, no universally accepted or fully established unified theory for iron-based superconductors exists.

Our theoretical analysis indicates that solely considering the first-order AFM fluctuation may not suffice to simulate the complete superconducting phase diagram of BaFe$_2$As$_2$. Antiferromagnetic fluctuations typically diminish under pressure. The pressure-dependent theoretical $T_c$ (second-order AFM) in BaFe$_2$As$_2$ is significantly enhanced, as the first-order AFM fluctuation does not decrease as rapidly as the second-order AFM fluctuation at high pressures. This aligns with the results of Coh et al[14] when a higher-order AFM fluctuation is employed. Both first and second-order AFM fluctuations can yield accurate theoretical $T_c$ values at low pressures, as it is feasible to fit a dependent variable (Y) linearly within the high-order term when the independent variable (X) is near zero or small. A similar scenario is observed in SrFe$_2$As$_2$, where only the second-order AFM fluctuations can accurately calculate $T_c$ at high pressure (the exchange factor drops by ~25% rapidly). However, for CaFe$_2$As$_2$, the first-order AFM fluctuation maintains precision in calculating $T_c$, as its exchange factor only decreases by less than 4% from 0.1GPa to 1.2GPa.

Elevating the Debye frequency permits higher energy phonons to interact with a greater number of electrons within the ARPES range of spectral weight shift[16-19]. This interaction results in an augmentation of the $R_{ARPES}$ value, as the effective electronic DOS increases. However, the $f(E_{ex})$ effects are typically counterbalanced by high pressure, which mitigates the impact of $R_{AF}$ and $R_{ph}$. The implementation of the second-order AFM fluctuation is crucial to decrease the $T_c$ of BaFe$_2$As$_2$ above ~2GPa. The $R_{ARPES}$ factor accounts for electron energies situated down to roughly 30meV (equivalent to ~350K) below the Fermi level. This method continues to adhere to the hyperbolic tangent shape of the Fermi-Dirac statistics across the Fermi level at a finite temperature[13]. The correlation between $T_c$ and AFM fluctuations, as depicted in Fig 1b and Fig 2b, underscores the significance of AFM fluctuations in the superconducting pairing process.

The reported experimental $T_c$ of Ba$_{0.8}$K$_{0.2}$Fe$_2$As$_2$ by Rotter et al.[32] and Böhmer et al.[8] are 26K and 20K, respectively. The slight discrepancy in the experimental $T_c$ values may be attributed to the experimental methodologies employed or the uncertainty in the tetrahedral Fe-As-Fe angle.

Utilizing the tetrahedral angle of BaFe$_2$As$_2$ measured by R. Mittal et al.[34] at low temperatures as a reference, the theoretical $T_c$ of Ba$_{0.8}$K$_{0.2}$Fe$_2$As$_2$ reverts to ~20K. However, we employ data from Rotter et al., as it offers a more systematic $T_c$ calculation, where all lattice parameters, bond lengths, and bond angles are derived from a single experimental setup[32]. While our theoretical $T_c$ values of Ba$_{1-x}$K$_x$Fe$_2$As$_2$ are not devoid of errors, the $T_c$ profiles depicted in Fig 2a align more closely with the experimental data. The dopant-induced pressure in Ba$_{1-x}$K$_x$Fe$_2$As$_2$ is relatively weak, hence the combined model based on first-order AFM fluctuation remains accurate. The same pairing strength formulae are applied for BaFe$_2$As$_2$, Ba$_{1-x}$K$_x$Fe$_2$As$_2$, CaFe$_2$As$_2$, SrFe$_2$As$_2$, MgFe$_2$As$_2$ with reasonably well-calculated $T_c$, suggesting that the pairing mechanism of these four undoped iron-based superconductors may share similar components from a statistical perspective. No $T_c$ has been reported for MgFe$_2$As$_2$; this may be due to the fact that as the $f(E_{ex})$ drops to zero above 3GPa, the pairing strength becomes zero and we propose that pressure is not an effective method to induce iron-based superconductivity in MgFe$_2$As$_2$.

The factors contributing to electron participation within the ARPES range of spectral weight shift in IBSC remain ambiguous, potentially involving a myriad of complex elements such as spin-orbital coupling, nematicity, antiferromagnetism, the competition between s-wave and d-wave pairing, and electron-phonon interaction[36-41]. In certain iron-based superconductors, the Fermi surface displays nematic order[41], which influences the superconducting order parameter. The precise effect of nematic order on the pairing symmetry on the Fermi surface is still an open question. However, the intricate interactions between electrons on the Fermi surface can be simplified by considering all high-energy electrons within the ARPES range of spectral weight shift. Here, high-energy electrons are not anticipated to involve either nematicity or 4-fold symmetry in the superconducting gap. The electron-phonon coupling on the Fermi surface typically weakens by ~20-30% when transitioning to 4-fold or d-wave like symmetry[42], which validates our calculated value of $\langle f_{angular} \rangle \sim 0.8$. In our $T_c$ calculation, we considered the 4-fold pairing symmetry across the entire ARPES range of spectral weight shift below the Fermi level. However, the change in $T_c$ is only a few Kelvin, as depicted in Figure 3. We have not examined the case of $\langle f_{angular} \rangle \leq 0.5$ because reducing the pairing strength by gap anisotropy necessitates the formation of a p-wave order parameter[43], and none of our studied compounds are expected to be p-wave superconductors.

Another potential source of error in $T_c$ could be the approximation of the Debye energy in the ARPES factor. A trend can be observed between the ARPES range and the Debye energy when comparing the ARPES data of different materials. An examination of the ARPES data of LiFeAs, FeSe, and FeSe/SrTiO$_3$ provides insights into this observation. In the case of FeSe/SrTiO$_3$, the $T_c$ is approximately 100K[16], and the interfacial phonon energy is around 1200K[44]. Concurrently, the ARPES range in this context is measured to be between 0.1-0.3eV[16]. Conversely, for LiFeAs or FeSe, the $T_c$ is lower[12], around 10K, and the phonon energy is approximately 300K. Correspondingly, the ARPES ranges for LiFeAs and FeSe are much narrower, around 0.03-0.06eV.

These data suggest a trend where an increase in Debye energy corresponds to an increase in the ARPES range, indicating a broader energy range for electronic excitations. Therefore, we select the Debye energy as an approximated energy range in the ARPES factor, but it may not precisely correspond to the actual energy range below the Fermi level. This approximation can lead to either an overestimation or underestimation of $T_c$. To rectify this issue, one could determine the exact ARPES range for each IBSC and use that value instead of an approximation. This approach would yield a more accurate estimation of the ARPES factor. However, scanning all ARPES data for every discovered IBSC would indeed require significant experimental effort and pose practical challenges and limitations.

Upon initial observation, the anisotropy factor associated with nematicity appears to exert minimal influence on the calculated $T_c$ values in Figure 3. However, this does not suggest that they are inconsequential to the mechanism of IBSC. In fact, they may play a pivotal role in initiating IBSC[45]. Once iron-based superconductivity is triggered, the impact of the anisotropy factor diminishes due to the renormalization of the strong pairing strength in compressed IBSC. Therefore, it would be unjust to assess the significance of the anisotropy factor or nematicity based solely on Figure 3. Conversely, the combined model may not be intended for discovering new IBSC compounds. Rather, it may be suitable for predicting $T_c$ at higher pressures or heavy doping levels if superconductivity has already been confirmed at ambient pressure, low pressure, or low doping concentration. Despite bridging the gap between theoretical and experimental $T_c$ values, the combined model does not explicitly provide a definitive statement about the pairing mechanism of iron-based superconductors. The unified theory of iron-based superconductors remains an open question necessitating further investigation and research. The possibility of other models capable of producing accurate theoretical $T_c$ of iron-based superconductors via entirely different mechanisms is not excluded.

**Conclusions**

We have successfully established a framework for simulating the superconducting phase diagrams of major 122-type intercalated IBSC with commendable accuracy. Our observations indicate that certain IBSC require the consideration of higher-order antiferromagnetic fluctuations, particularly under high-pressure conditions. This is a pivotal step towards integrating various effects into the pairing mechanism of IBSC, including superconducting gap anisotropy, spin-orbital coupling, high-energy electrons, abnormal phonon behavior, screening effects, spin density wave, and antiferromagnetism. Our findings pave the way for a deeper understanding of the complex interplay of factors governing superconductivity in these materials.

**Data Availability Statement**

Data are sharable under reasonable request. The authors are usually supportive for reproducing the results if further assistance is needed (please send your technical requests to roywch654321@gmail.com).

**Acknowledgments**

We thanks the Department of Industrial and Systems Engineering in The Hong Kong Polytechnic University to provide simulation support.

**References**

[1] S. Sharma, S. Bharathia, S. Chandra, V. R. Reddy, S. Paulraj, A. T. Satya, V. S. Sastry, A. Gupta, C. S. Sundar, Superconductivity in Ru-substituted polycrystalline BaFe$_{2-x}$Ru$_x$As 2, *Phys. Rev. B* **81**, 174512 (2010).

[2] S. Kasahara, T. Shibauchi, K. Hashimoto, K. Ikada, S. Tonegawa, R. Okazaki, H. Shishido, H. Ikeda, H. Takeya, K. Hirata, T. Terashima, Y. Matsuda, Evolution from non-Fermi- to Fermi-liquid transport via isovalent doping in BaFe$_2$(As$_{1-x}$P$_x$)$_2$ superconductors, *Phys. Rev. B* **81**, 184519 (2010).

[3] M. Rotter, M. Tegel, D. Johrendt, Superconductivity at 38 K in the Iron Arsenide (Ba$_{1-x}$K$_x$)Fe$_2$As$_2$, *Phys. Rev. Lett.* **101**, 107006 (2008).

[4] A. S. Sefat, R. Jin, M. A. McGuire, B. C. Sales, D. J. Singh, D. Mandrus, Superconductivity at 22 K in Co-Doped BaFe$_2$As$_2$ Crystals, *Phys. Rev. Lett.* **101**, 117004 (2008).

[5] J. Paglione, R. L. Greene, High-temperature superconductivity in iron-based materials, *Nat. Phys.* **6**, 645 – 658 (2010).

[6] J. M. Allred, K. M. Taddei, D. E. Bugaris, M. J. Krogstad, S. H. Lapidus, D. Y. Chung, H. Claus, M. G. Kanatzidis, D. E. Brown, J. Kang, R. M. Fernandes, I. Eremin, S. Rosenkranz, O. Chmaissem, R. Osborn, Double-Q spin-density wave in iron arsenide superconductors, *Nat. Phys.* **12**, 493–498 (2016).

[7] J. Hou, C.-w. Cho, J. Shen, P. M. Tam, I-H. Kao, M. H. G. Lee, P. Adelmann, T. Wolf, R. Lortz , Possible coexistence of double-Q magnetic order and chequerboard charge order in the re-entrant tetragonal phase of Ba$_{0.76}$K$_{0.24}$Fe$_2$As$_2$, *Physica C* **539**, 30–34 (2017).

[8] A. E. Böhmer, F. Hardy, L. Wang, T. Wolf, P. Schweiss, C. Meingast, Superconductivity-induced re-entrance of the orthorhombic distortion in Ba$_{1-x}$K$_x$Fe$_2$As$_2$, *Nat. Commun.* **6**:7911(2015).

[9] Patricia L Alireza, Y T Chris Ko, Jack Gillett, Chiara M Petrone, Jacqueline M Cole, Gilbert G Lonzarich and Suchitra E Sebastian, Superconductivity up to 29 K in SrFe2As2 and BaFe2As2 at high pressures, J. Phys.: Condens. Matter 21, 012208 (2009)

[10] Y Zheng, Y Wang, B Lv, C W Chu and R Lortz, Thermodynamic evidence for pressure-induced bulk superconductivity in the Fe–As pnictide superconductor CaFe2As2, New Journal of Physics 14 053034 (2012)

[11] A. E. Böhmer, A. Kreisel, Nematicity, magnetism and superconductivity in FeSe, *J. Phys.: Condens. Mat.* **30**, 023001 (2018).

[12] B. Li, Z. W. Xing, G. Q. Huang, M. Liu, Magnetic-enhanced electron-phonon coupling and vacancy effect in "111"-type iron pnictides from first-principle calculations, J. App. Phys. 111, 033922 (2012).

[13] S. Deng, J. Köhler, A. Simon, Electronic structure and lattice dynamics of NaFeAs, Phys. Rev. B 80, 214508 (2009).

[14] S. Coh, M. L. Cohen, S. G. Louie, Anti-ferromagnetism enables electron-phonon coupling in iron-based superconductors, Phys. Rev. B 94, 104505 (2016).

[15] S.-F. Wu, W.-L. Zhang, V. K. Thorsmølle, G. F. Chen, G. T. Tan, P. C. Dai, Y. G. Shi, C. Q. Jin, T. Shibauchi, S. Kasahara, Y. Matsuda, A. S. Sefat, H. Ding, P. Richard, and G. Blumberg, Physical Review Research, 2, 033140 (2020).

[16] X.-W. Jia, H.-Y. Liu, W.-T. Zhang, L. Zhao *et al.*, Common Features in Electronic Structure of the Oxypnictide Superconductors from Photoemission Spectroscopy, *Chin. Phys. Lett.* **25**, 3765-3768 (2008).

[17] C. Liu, G. D. Samolyuk, Y. Lee, N. Ni, T. Kondo, A. F. Santander-Syro, S. L. Bud'ko, J. L. McChesney, E. Rotenberg, T. Valla, A. V. Fedorov, P. C. Canfield, B. N. Harmon, A. Kaminski, K-doping dependence of the Fermi surface of the iron-arsenic $Ba_{1-x}K_xFe_2As_2$ superconductor using angle-resolved photoemission spectroscopy, *Phys Rev Lett.* **101**, 17 (2008)

[18] C. Zhang *et al.*, Ubiquitous strong electron–phonon coupling at the interface of $FeSe/SrTiO_3$, *Nat. Commun.* **8**, 14468 (2017).

[19] U. Stockert, M. Abdel-Hafiez, D. V. Evtushinsky, V. B. Zabolotnyy, A. U. B. Wolter, S. Wurmehl, I. Morozov, R. Klingeler, S. V. Borisenko, B. Büchner, Specific heat and angle-resolved photoemission spectroscopy study of the superconducting gaps in LiFeAs, *Phys. Rev. B* **83**, 224512 (2011).

[20] A. Mani, N. Ghosh, S. Paulraj, A. Bharathi, C. S. Sundar, Pressure-induced superconductivity in $BaFe_2As_2$ single crystal, *Europhys. Lett.* **87**, 17004 (2009).

[21] F. Hardy, A. E. Böhmer, D. Aoki, P. Burger, T. Wolf, P. Schweiss, R. Heid, P. Adelmann, Y. X. Yao, G. Kotliar, J. Schmalian and C. Meingast, Evidence of Strong Correlations and Coherence-Incoherence Crossover in the Iron Pnictide Superconductor $KFe_2As_2$, *Phys. Rev. Lett* **111**, 027002 (2013).

[22] Y.-f. Yang, D. Pines, Emergent states in heavy-electron materials, *PNAS* **109**, E3060–E3066 (2012).

[23] V. Grinenko, P. Materne, R. Sarkar, H. Luetkens, K. Kihou, C. H. Lee, S. Akhmadaliev, D. V. Efremov, S.-L. Drechsler, H.-H. Klauss, Superconductivity with broken time-reversal symmetry in ion-irradiated $Ba_{0.27}K_{0.73}Fe_2As_2$ single crystals, *Phys. Rev. B* **95**, 214511 (2017).

[24] J. Richard Christman, Fundamentals of Solid States Physics, Wiley (1986).

[25] H. Miao, T. Qian, X. Shi, P. Richard, T. K. Kim, M. Hoesch, L. Y. Xing, X.-C. Wang, C.-Q. Jin, J.-P. Hu & H. Ding, Observation of strong electron pairing on bands without Fermi surfaces in $LiFe_{1-x}Co_xAs$, Nature Communications volume 6, Article number: 6056 (2015)

[26] S. V. Borisenko, D. V. Evtushinsky, Z.-H. Liu, I. Morozov, R. Kappenberger, S. Wurmehl, B. Büchner, A. N. Yaresko, T. K. Kim, M. Hoesch, T. Wolf & N. D. Zhigadlo, Nature Physics, Volume 12, pp 311–317 (2016)

[27] W. L. McMillian, Transition Temperature of Strong-Coupled Superconductors, *Phys. Rev.* **167**, 331 (1968).

[28] J. P. Perdew, *et al.*, Atoms, molecules, solids, and surfaces: Applications of the generalized gradient approximation for exchange and correlation, *Phys. Rev. B* **46**, 6671 (1992).

[29] A. D. Becke, Density-functional exchange-energy approximation with correct asymptotic behavior, *Phys. Rev. A* **38**, 3098 (1988).

[30] D.J.Kim, The influence of magnetism on the electron-phonon interaction in metals, Physica 91B, pp 281-287 (1977)

[31] Y. Wen, D. Wu, Re. Cao, L. Liu, L. Song, The Third-Order Elastic Moduli and Debye Temperature of SrFe2As2 and BaFe2As2: a First-Principles Study, J. Supercond. Nov. Magn. 30, 1749–1756 (2017).

[32] M. Rotter, M. Panger, M. Tegel, D. Johrendt, Superconductivity and Crystal Structures of $(Ba_{1-x}K_x)Fe_2As_2$ (x = 0 - 1), *Angew. Chem. Int. Ed.* **47**, 7949 –7952 (2008).

[33] H. Oh, S. Coh, M. L. Cohen, Calculation of the specific heat of optimally K-doped $BaFe_2As_2$, J. Phys.: Condens. Matter 27, 335504 (2015)

[34] R. Mittal, S. K. Mishra, S. L. Chaplot, S. V. Ovsyannikov, E. Greenberg, D. M. Trots, L. Dubrovinsky, Y. Su, Th. Brueckel, S. Matsuishi, H. Hosono, G. Garbarino, Ambient- and low-temperature synchrotron x-ray diffraction study of $BaFe_2As_2$ and $CaFe_2As_2$ at high pressures up to 56 GPa, *Phys. Rev. B* **83**, 054503 (2011).

[35] J. R. Jeffries, N. P. Butch, K. Kirshenbaum, S. R. Saha, S. T. Weir, Y. K. Vohra, and J. Paglione, The suppression of magnetism and the development of superconductivity within the collapsed tetragonal phase of Ca0.67Sr0.33Fe2As2 at high pressure, Phys. Rev. B 85, 184501 (2012)

[36] J.-Ph. Reid, M. A. Tanatar, A. Juneau-Fecteau, R. T. Gordon, S. Rene´ de Cotret, N. Doiron-Leyraud, T. Saito, H. Fukazawa, Y. Kohori, K. Kihou, C. H. Lee, A. Iyo, H. Eisaki, R. Prozorov, L. Taillefer, Universal Heat Conduction in the Iron Arsenide Superconductor $KFe_2As_2$: Evidence of a d-Wave State, *Phys. Rev. Lett.* **109**, 087001 (2012).

[37] F. F. Tafti, A. Juneau-Fecteau, M.-È. Delage, S. René de Cotret, J.-Ph. Reid, A. F. Wang, X.-G. Luo, X. H. Chen, N. Doiron-Leyraud, L. Taillefer, Sudden reversal in the pressure dependence of $T_c$ in the iron-based superconductor $KFe_2As_2$, *Nat. Phys.* **9**, 349–352 (2013).

[38] Ptok, A.; Rodríguez, K.; Kapcia, K.J. Superconducting monolayer deposited on substrate: Effects of the spin-orbit coupling induced by proximity effects. Phys. Rev. Mater. 2, 024801 (2018).

[39] Hutchinson, J.; Hirsch, J.E.; Marsiglio, F. Enhancement of superconducting Tc due to the spin-orbit interaction. Phys. Rev. B, 97, 184513. (2018)

[40] Qisi Wang, Lara Fanfarillo and Anna E. Böhmer et al, Nematicity in iron-based superconductors, Frontiers in Physics Condensed Matter Physics, Volume 10, 1-120 (2022)

[41] Wiecki P, Rana K, Böhmer AE, Lee Y, Bud'ko SL, Canfield PC, et al. Persistent Correlation between Superconductivity and Antiferromagnetic Fluctuations Near a Nematic Quantum Critical Point in FeSe1−xSx. Phys Rev B (2018) 98:020507. doi:10.1103/PhysRevB.98.020507

[42] J. Song, J. F. Annett, Electron-phonon coupling and d-wave superconductivity in the cuprates, Phys. Rev. B 51, 3840 (1995).

[43] Ke-Chuan Weng & C. D. Hu, The p-wave superconductivity in the presence of Rashba interaction in 2DEG, Scientific Reports, Volume 6, Article number: 29919 (2016)

[44] Zhang, S.; Guan, J.; Wang, Y.; Berlijn, T.; Johnston, S.; Jia, X.; Liu, B.; Zhu, Q.; An, Q.; Xue, S.; et al. Lattice dynamics of ultrathin FeSe films on SrTiO3. Phys. Rev. B 2018, 97, 035408

[45] Qisi Wang, Lara Fanfarillo, Anna E. Böhmer, Editorial: Nematicity in iron-based superconductors, Front. Phys, Sec. Condensed Matter Physics (2022)